\documentclass[]{raa}            
\usepackage{graphicx,times}
\usepackage{natbib}
\usepackage{amssymb,amsmath}

\usepackage[pagebackref=true]{hyperref}
\hypersetup{pdftitle = Environmental Dependency of Galaxies in EAGLE, pdfauthor = Mingge Zhang, pdfsubject= Astronomy, pdfkeywords = Galaxy alignment} 
\hypersetup{colorlinks = true, linkcolor = green, anchorcolor = red, citecolor = blue, filecolor = red, pagecolor = red, urlcolor = red}

\bibpunct{(}{)}{;}{a}{}{,}

\graphicspath{{./}{figures/}}
\newcommand{\reffig}[1]{Figure \ref{#1}}
\newcommand{\reftab}[1]{Table \ref{#1}}

\begin{document}

\title{Alignment between Satellite and Central Galaxies in the Eagle Simulation: Dependence on the Large-Scale Environments
}

\volnopage{Vol.0 (20xx) No.0, 000--000}      
\setcounter{page}{1}          

\author{Mingge Zhang
	\inst{1}
	\and Yang Wang$^\dag$
	\inst{1}
}

\institute{School of Physics and Astronomy, Sun Yat-sen University, Zhuhai 519082, China; {\it wangyang23@mail.sysu.edu.cn}\\
\vs\no
{\small Received~~2019 month day; accepted~~2019~~month day}
}

\abstract{
	The alignment between satellite and central galaxies serves as a proxy for
	addressing the issue of galaxy formation and evolution and has been
	investigated abundantly in observations and theoretical works. Most scenarios
	indicate that the satellites preferentially locate along the major axis of their
	central galaxy. Recent work shows that the strength of alignment signals depends
	on large-scale environment in observations. We use the publicly-released data from EAGLE to figure out whether the
	same effect can be found in the hydrodynamic simulation. We found much stronger
	environmental dependency of alignment signal in simulation. And we also explore
	change of alignments to address the formation of this effects.
}
\authorrunning{M.G. Zhang \& Y. Wang}            
\titlerunning{Environmental Dependency of Galaxy Alignments}  

\maketitle

\section{Introduction}

The standard $\Lambda$ cold dark matter cosmology suggests a hierarchical
scenario of cosmic structure formation. The growth of Gaussian
density fluctuations via highly nonlinear and anisotropic
gravitational clustering process shape the
large scale structure of universe with four distinct environments, i.e.
cluster, filament, sheet and void \citep{Joeveer1978,Bond1996}.
Flowing out of the voids,  matter accrete on the sheets, then collapse onto the
filaments and finally assemble to form clusters at the intersections of
filaments. On smaller scales, dark matter collapses to small
halos firstly and then may go through merges with other halos to form larger halo or been
captured by larger halos to become their "sub-halos", and galaxies are formed in
the inner region of these dark matter halos (\citealt{White1978}). According to this
paradigm, galaxies are not
randomly distributed in the universe, but rather show some alignments, i.e.
the shape, position , spin etc. tend to have preferential
directions (\citealt{Jing2002, Aubert2004, Bagla2006,Aragon-Calvo2010, Arieli2010}). Thus, the galaxy alignment could
be an indicator for probing the galaxy formation and evolutionary history in the $\Lambda$ CDM universe.

Many observational and theoretical works have confirmed galaxy alignments.
Explorations on galaxy alignments in observation started began with the early works \cite{Sastry1968}
and \cite{Holmberg1969}. The former reported an alignment between satellite
galaxies and major axis of central galaxies. But the latter claimed the opposite
results that the position vectors of satellite galaxies tend to be perpendicular
to the major axis of central galaxies. The disagreement of these two works
may partly result from the small survey volume of the galaxy sample, such a defect has been overcome by the
emergence of large galaxy surveys, e.g., Sloan Digital Sky Survey (SDSS).
Benefited from super-large volume of galaxy sample, studies on central-satellite galaxy alignments
has already reached a unified and largely accepted conclusion that the satellites tend to
align with the major axis of central galaxy (e.g., \citealt{Agustsson2006b, Yang2006,Brainerd2005,Agustsson2010}).
Following theoretical works by utilizing semi-analytical models, both N-body simulations
and hydrodynamical simulations has confirmed such a trend (\citealt{Agustsson2006a, Kang2007, Libeskind2005, Libeskind2007,Codis2012,Codis2015}). For a full overview on all kinds of issues about galaxy alignments, in which the central-satellite alignment is just one aspect, readers could refer
to \cite{Schaefer2009d,Joachimi2015,Kiessling2015,Kirk2015}.


The central-satellite alignment comes from the combination of smooth mass
accretion and mergers of dark matter halos. The anisotropic collapse of dark
matter halo will shape its central galaxy with preferential directions
\cite{Jing2002, Schaefer2009d}. As a result, the direction of axes, or
angular momentum, of central galaxies
will relate with their surrounding structures.(\citealt{Zhang2013,Zhang2015})
On the other hand, as remnants of accreted
halos, merger events can be inferred from the position of satellite galaxies.
Thereby the alignment between the position of satellite galaxies and direction of large scale structures was also
studied (e.g., \citealt{Tempel2015a}). However,
detecting these two processes separately could be tough due to sort of ambiguities in
defining the shape and direction of large scale structures from observational data,
leading to most work focusing on the central-satellite alignments.

In the different structure type of the cosmic web, e.g., cluster, filament, sheet and void,
the central-satellite alignment could be quite different, because of either dark matters collapsing via different directions
(\citealt{Codis2018a}), or subhaloes (satellite galaxies) being accreted via different paths dictated by large scale structures \cite{Libeskind2014}.
Accordingly, the shape of central galaxy and the distribution
of satellite galaxies may be influenced by the local structure types.
Many works found that the shape of dark matter halo, which should be shapely aligned with its central
galaxy, is related with local structures (\citealt{Hahn2007, zhang2009, Forero2014}). \cite{Tempel2013} found that the
minor axes of elliptical galaxies are preferentially perpendicular to hosting
filaments but the alignment signal is weak in sheets. And the spin axes of spirals align with host
filaments, but there is no alignment signal between the spiral spin and the sheet normal vector.
\cite{Codis2018a} confirmed the same trend and further revealed an alignment
flipping phenomenon from high to low redshifts. Some works like \cite{Tempel2015a} confirmed that the angular positions
of satellite galaxies tend to align with filaments. A further
explanation is provided by \cite{Libeskind2015}, in which they claimed that
the plane of satellite galaxies' orbit is well aligned with the
collapse direction derived from the shear tensor of environmental
velocity fields.

Since the large scale structures have influences on both central galaxies
and satellite galaxies, we would expect the alignment between them depending on the local cosmic environments.
It would be interesting to examine this issue, and recently \cite{Wang2018a} have done such exploration.
\cite{Wang2018a} found environment dependence of the alignment between satellite and central galaxies in the SDSS DR7 data.
Following \cite{Wang2018a}, we would like to examine  whether this dependence exists for simulated galaxies in
cosmological hydrodynamical simulation, and to explore the possible explanation for this phenomenon.

The paper is organized as follows: we brief firstly the simulation
and the galaxy catalogue, definitions of large-scale structure and alignment
angles in Section~\ref{sec:method}. Then the main results are given
in Section~\ref{sec:result}. Finally, we give discussions and conclusions
in Section~\ref{sec:con}.

\section{Data and Methodology}
\label{sec:method}

In this work, we use the publicly released data from the Evolution and
Assembly of Galaxies and their Environments (EAGLE) simulation
(\citealt{Schaye2015}) , which was run using a modified version of the code
GADGET 2 (\citealt{Springel2005}). The cosmology parameters are
$\Omega_{m}=0.307,\Omega_{b}=0.04825,\sigma_{8}=0.8288,n_{s}=0.9611, h=0.6777$.
We take advantage of the
simulation run labelled of  “Ref-
L100N1504”, i.e., with a box size of $100 \rm Mpc$, particles number of
$2\times 1504^{3}$ , and softening length of $2.66 \rm kpc$. The mass resolutions of gas and dark matter
particles are $1.81 \times 10^{6} h^{-1}M_{\odot}$ and $9.70 \times 10^{6}
	h^{-1}M_{\odot}$.
Stellar particles have variable mass around $7\times10^{5} h^{-1}M_{\odot}$.
More details about EAGLE simulation can be found
in \cite{Schaye2015}.

\subsection{Galaxies Samples}
We use five snapshots (z=0,1,2,3,5) from the simulation for data analysis. The
dark matter halos are identified by the standard friends-of-friends (FOF)
algorithm \citep{Davis1985}. Dark matter particles within a linking
length of $0.2$ times of the mean inter particle separation are assigned to the
same dark matter halo. Gas and star particles are assigned to the FoF halo
where their nearest dark matter particles reside in. The subhaloes are found by
the SUBFIND algorithm (\citealt{Springel2001, Dolag2009}). We select
central-satellite pairs from the $10000$ most massive dark matter halos.
In remaining less massive halos, no satellite galaxies could be found due to the
small number of star particles.
We further place a constrain on the galaxy sample that only galaxies with
more than $100$ star particles are taken for analysis, corresponding to stellar mass of $7\times10^{7}h^{-1}M_{\odot}$.
The number of central-satellite galaxy pairs selected is listed in Table \ref{Table1}.
\begin{table}[h]
	\centering
	\caption{Number of central-satellite pairs selected at different redshifts.}
	\begin{tabular}{ccccccc}
		\hline
		z           & 0     & 1     & 2     & 3    & 5    & \\
		\hline
		pair number & 21693 & 14566 & 10740 & 5914 & 1168 & \\
		\hline
	\end{tabular}
	\label{Table1}
\end{table}

\subsection{Characterize The Large Scale Environment}

\begin{figure}[hbtp]
	\centering
	\includegraphics[width=0.47\linewidth]{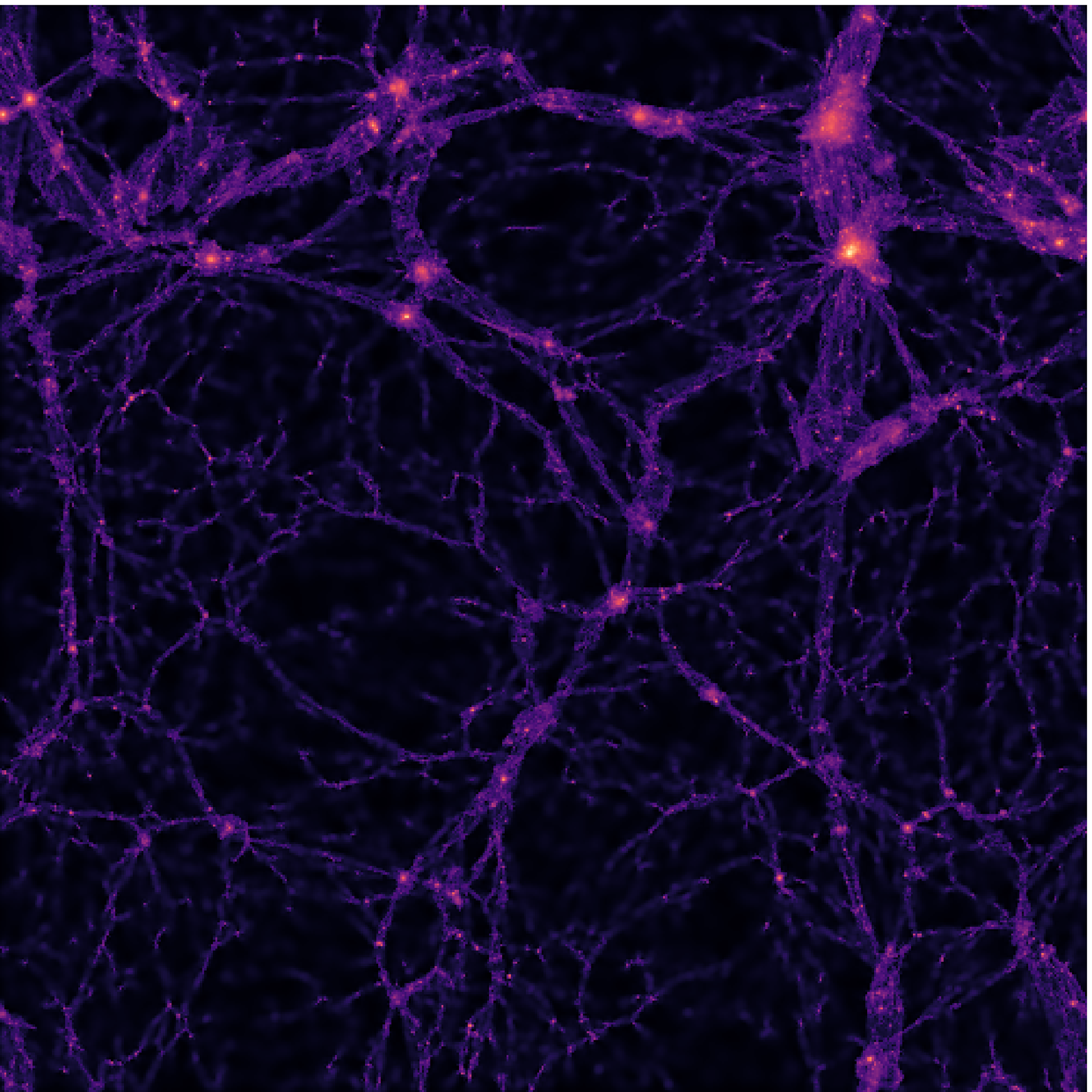}
	\quad
	\includegraphics[width=0.47\linewidth]{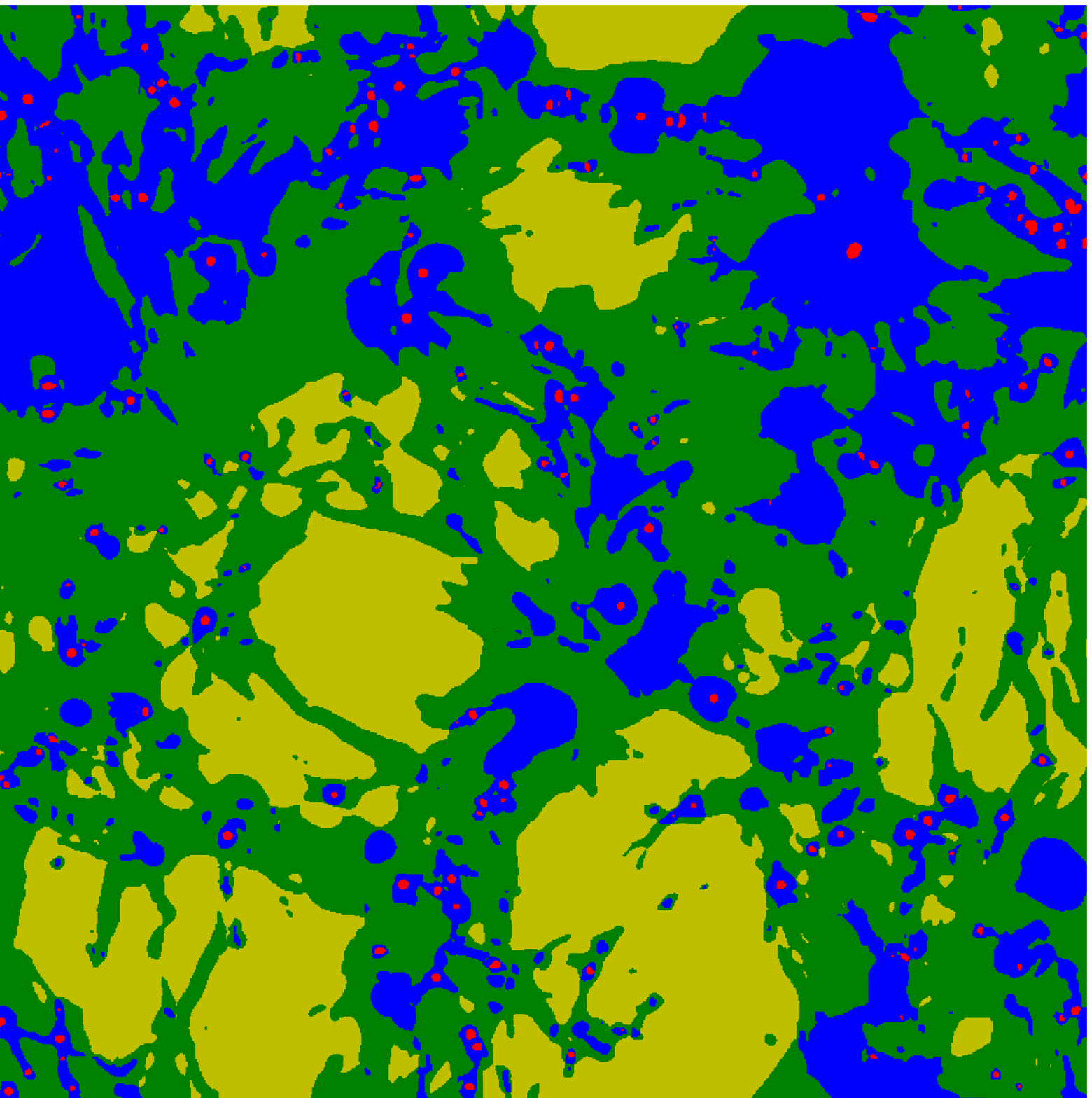}
	\caption{Left figure shows the density field of a slice with volume of $100{\rm Mpc} \times 100
				{\rm Mpc} \times 60 {\rm Kpc}$ at redshift $0$. Right Figure indicate the large scale structure within the same region. Red regions are clusters,
		greens are filaments, blues are sheets and yellows are voids.
	}
	\label{fig:fig1}
\end{figure}

The structure types in the large scale environment, namely cluster, filament, sheet and void, are
defined following the same method applied in \cite{Hahn2007}, \cite{Forero-Romero2009} and \cite{Zhu2017}.
We calculate the three eigenvalues of the Hessian matrix of the tidal field on specific grids
\begin{equation}
	T_{i,j} = \frac{\partial^{2}\phi}{\partial{r_{i}}\partial{r_{j}}}
\end{equation}
where $\phi$ is the potential there and $i,j$ go from $1$ to $3$ to cover the $3$ directions. For each grid,  we count the
numbers of eigenvalues above a certain threshold $\lambda_{t}$.
If all three eigenvalues are larger than $\lambda_t$, the structure type around that
point will be tagged cluster. Similarly,  we sort out all particle, each with two, one or none eigenvalues larger than the threshold $\lambda_t$
corresponding to filament, sheet and void respectively. In many works $\lambda_{t}$ is set to be $0$ (\citealt{Hahn2007}).
However, some works suggest a larger value of $\lambda_{t}$ (\citealt{Zhu2017, Forero-Romero2009}) to avoid producing smaller
volume of voids with $\lambda_t=0$ than the theoretical prediction.
Visually, a reasonable classification of the large scale structure can be given by setting $\lambda=2.0$.
\reffig{fig:fig1} displays the matter distribution(left) and corresponding environments(right) in a slice of simulation at redshift $0$, where cells of
different structures are assigned with different colors.

\subsection{Characterize The Alignment}

\begin{figure}[hbtp]
	\centering
	\includegraphics[width=1\linewidth]{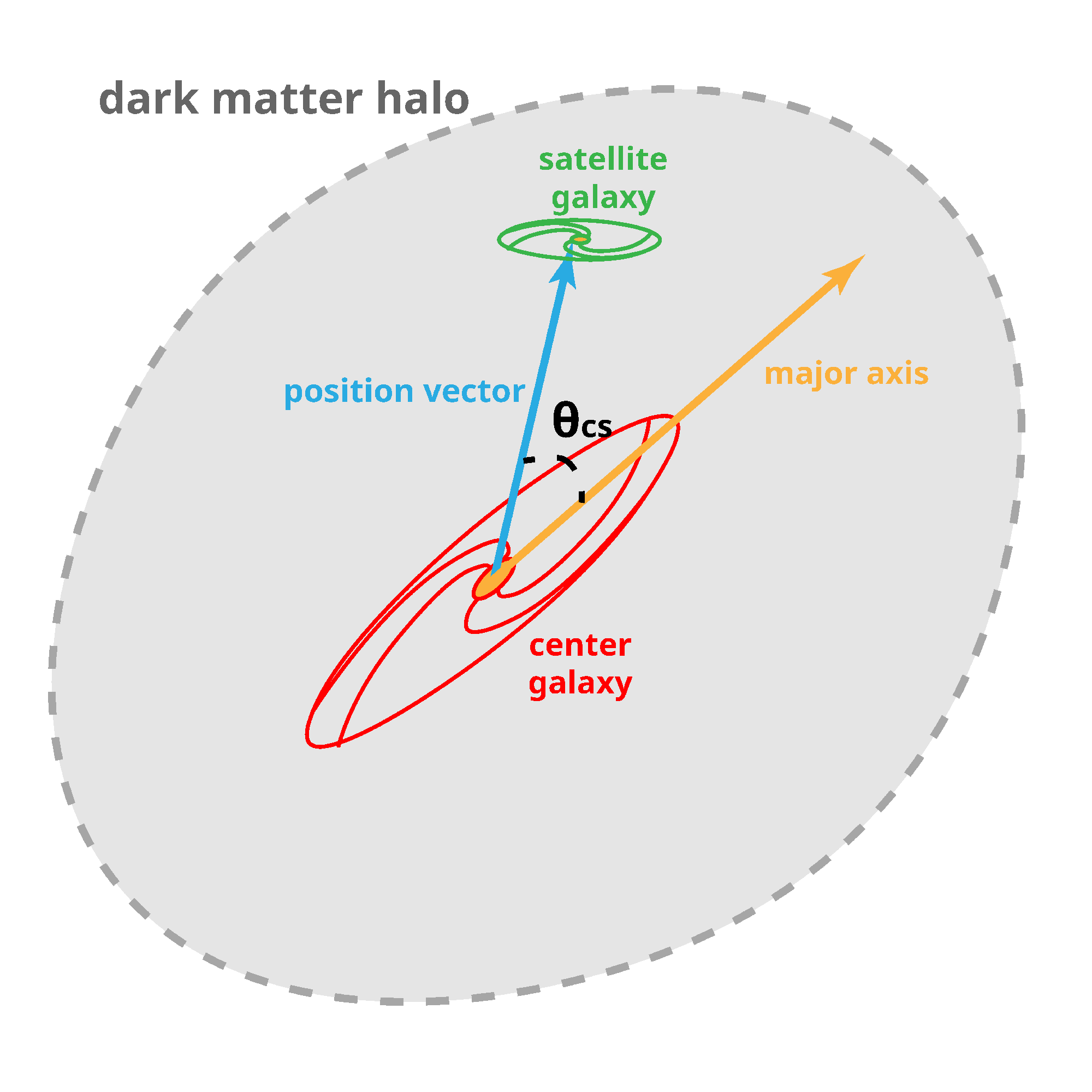}
	\caption{A figure illustrating the central-satellite alignment angle. The
		red and green ellipse represent the central and satellite galaxies. The
		yellow vector represents the major axis of center galaxy. The blue vector
		direct from the center of central galaxy to the center of satellite galaxy.
		The angular separation between blue and yellow vectors $\theta_{CS}$ is the
		alignment angle.}
	\label{fig2}
\end{figure}

In this paper we analyze the alignment angle between the major axis direction
of the central galaxy and the position direction of satellites, as illustrated
in \reffig{fig2}. The red arrow represents the major axis of the
central galaxy. The blue arrow represents the direction of the satellite
relative to the central of the central galaxy, and $\theta_{CS}$ is the
alignment angle. The major axis of the
central galaxy is determined by the mass weighted shape matrix whose element
is defined as:
\begin{equation}
	I_{ij}=\frac{\sum_{k}m_{k}x_{k,i}x_{k,j}}{\sum_{k}m_{k}}
\end{equation}
where $m_{k}$ is the mass of the $k\rm th$ star particle in the central galaxy,
, $x_{k,i}$ is the coordinate of $k \rm th$ star particle along $i \rm th$ axis ($i$ ranges from $1$ to $3$),
and the summation is taken over all the star particles in the central galaxy.  Once the shape matrix $I=\{I_{ij}, i,j=1,2,3\}$
is obtained, the major axis can be specified by the eigenvector corresponding to the maximum eigenvalue of the shape matrix.

The alignment angle $\theta_{CS}$  ranges from $0^{\circ}$ to $90^\circ$,
and a relative small value of $\theta<45^\circ$ implies preferentially aligned distribution along the major
axis of central galaxy. To compare with observations, we project central
galaxies and satellites onto a 2-dimensional plane before calculating the
alignment angle $\theta_{CS}$. The projection onto the $x-y$ plane can be made simply by setting the $z$ coordinate of all particles to be zero.
We also tested projections onto the $x-z$ and $y-z$ plane.  No significant differences have been found among the three projecting direction.
In the following, results are obtained by projected galaxies on the $x-y$ plane if not specified elsewhere.

\section{Results}
\label{sec:result}
In this section, we first present the color and mass dependence of
central-satellite alignment following the previous works \cite{Yang2006, Wang2018a}.
Then we check numerically the dependence on various structure types in the cosmic web to test whether we could reproduce
\cite{Wang2018a}'s results. Finally, we make an attempt to explore the origin of dependency on large scale environment by tracing alignment signal $\theta_{CS}$ through the cosmic time.

\subsection{The color and mass dependence of central-satellite alignments}
\label{subsec:ot}

Following \cite{Yang2006} and \cite{Wang2018}, we first check the overall
alignment signal. As shown in the first row of \reftab{Table2}, the mean alignment
angle in EAGLE simulation is smaller than \cite{Yang2006} and
\cite{Wang2018}, indicating a stronger alignment. According to the previous works
e.g. \cite{Kang2007, Faltenbacher2009}, the alignment between subhalo and the
major axis of host halo is much stronger than that between satellite
galaxies and the major axis of central galaxies. To reproduce the alignment
signal inferred from the observations, the misalignment between central galaxy and host halo is
necessary inevitably. \cite{Wang2014b,Faltenbacher2009} suggest a misalignment angle
around $30^{\circ}$, which has been justified by \cite{Dong2014} using the hydrodynamical
cosmological simulation. However, we calculate the misalignment between
central galaxy and host halo in the EAGLE's data, and found that it peaks at about $20^{\circ}$.
This could be a reason why the stronger alignment signal has been found in that work. The
physics behind such small misalignment might be complex, thus we leave it for future work.

\begin{table}[h]
	\centering
	\caption{Mean alignment angle $\theta_{CS}$ of different (sub-)samples of central-satellites
		galaxy pairs at redshift $0$ in EAGLE simulation (fourth column),
		compared with the results in \cite{Yang2006}(second column) and \cite{Wang2018a}(third column). Sub
		samples are divided according to galaxy color. In row $2$ to $3$ only the
		color of central galaxies is considered. In row $4$ and $5$, we only consider
		the satellites' color. In row $6$ to $9$, the sample make constrains on both
		centrals and satellites. The sample name 'red - blue' means the galaxy paris have red centrals and blue satellites.}
	\begin{tabular}{cccc}
		\hline
		Sample Name     & Y06                  & W18                   & This Work            \\
		\hline
		all sample      & $42.2\pm0.2^{\circ}$ & $42.2\pm0.06^{\circ}$ & $38.1\pm0.3^{\circ}$ \\
		\hline
		red centrals    & $41.5\pm0.2^{\circ}$ & $41.7\pm0.10^{\circ}$ & $38.2\pm0.3^{\circ}$ \\
		blue centrals   & $44.5\pm0.5^{\circ}$ & $44.7\pm0.15^{\circ}$ & $38.0\pm0.7^{\circ}$ \\
		\hline
		red satellites  & $41.5\pm0.3^{\circ}$ & $41.5\pm0.11^{\circ}$ & $36.0\pm0.5^{\circ}$ \\
		blue satellites & $43.3\pm0.3^{\circ}$ & $43.2\pm0.09^{\circ}$ & $39.3\pm0.3^{\circ}$ \\
		\hline
		red - red       & $40.8\pm0.3^{\circ}$ & $40.9\pm0.12^{\circ}$ & $36.0\pm0.5^{\circ}$ \\
		red - blue      & $42.9\pm0.3^{\circ}$ & $42.6\pm0.13^{\circ}$ & $39.5\pm0.4^{\circ}$ \\
		blue - red      & $44.8\pm0.7^{\circ}$ & $45.5\pm0.31^{\circ}$ & $36.6\pm1.3^{\circ}$ \\
		blue - blue     & $44.2\pm0.6^{\circ}$ & $44.4\pm0.20^{\circ}$ & $38.4\pm0.8^{\circ}$ \\
		\hline
	\end{tabular}
	\label{Table2}
\end{table}

The rest rows of \reftab{Table1} show the color dependence of the alignment signal.
The color of a galaxy is defined by its magnitude $g-r$. For the probability
distribution of $g-r$,  there appear two peaks. We use the median value of these
two peaks to classify galaxies into the red and blue branches. This method was
suggested by \cite{Baldry2004b} and has been widely used.
In \cite{Yang2006} and \cite{Wang2018}, the red centrals or red satellites have
smaller alignment angles. Galaxies in EAGLE are unable to fully reproduce such trends.
In our samples, the red satellites align with the centrals' major axis more strongly, while for the red and blue central galaxies, they have
similar alignment signals.
However, it is always a challenge to fully recover the alignment dependency on the galaxy
color. Our results are quite similar to \cite{Dong2014}. Their results also
eliminate the differences between blue and red central galaxies. It is basically
caused by the fact that relatively more blue central galaxies are produced in the simulation than observation. As
displayed in figure 2 and 4 in \cite{Trayford2015}, the $g-r$ versus $M_*$ or $g-r$ versus $M_r$ profile in the EAGLE simulation
has similar outer shape and blue peak as GAMA galaxies, but it does not recover the red peak.
Thus some central galaxies with the strong alignment are mis-assigned to blue color.
\cite{Trayford2015} suggested that the flat slope of red sequence may be
attributed to the rapidly decreased stellar metallicity at low mass range,
and such decrement may be a resolution issue of the simulation.

\begin{figure}[hbtp]
	\centering
	\includegraphics[scale=0.4]{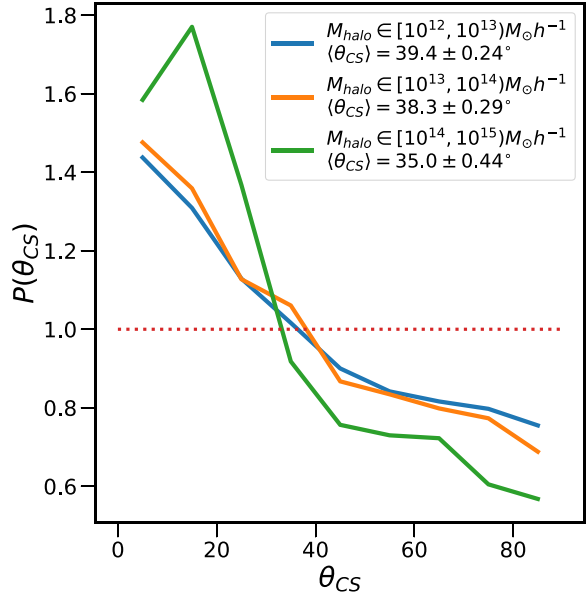}
	\caption{The normalized probability distribution of the alignment
		angle $\theta_{CS}$. Central-satellite pairs are assigned into $3$
		sub samples according to their host halo's mass, as indicated by
		the legend. The mean angle for each sub sample is also shown in the
		legend. The red dotted horizontal line $P=1$ represents a ideal isotropic
		distribution.}
	\label{angle_mass}
\end{figure}

We also check the dependence on the mass of host halo. The probability distribution
of the alignment angle $\theta_{CS}$ is shown in \reffig{angle_mass}. The
probability distribution is obtain by computing $P(\theta)=N(\theta)/N_R(\theta)$.
$N(\theta)$ is the number of central-satellite pairs with the alignment angle of
$\theta$ in the samples to be studied. $N_R(\theta)$ is the number of galaxy pairs with
the same $\theta$ in a sample with randomly distributed satellite galaxies.
The total number of galaxy pairs of random sample is the same as the simulation
sample. \reffig{angle_mass} shows that the alignment is clearly dependent
on the halo mass. Galaxy pairs in the massive halos exhibit stronger alignment.
This trends are in agreement with \cite{Yang2006,Kang2007, Wang2014b, Wang2018}.

In summary, the EAGLE's galaxy catalogue reproduces both the mass and color
dependence (partly) of alignment in the observations. It also encounters problems like
too stronger alignment signal and blue biased central galaxies, but these problems also exist in many other simulations.
On the other hand, the overall trend is quite reasonable. Thus the drawback does not border the main part of our
work, exploring the environmental effect of large scale structures on the alignment.

\subsection{The Environment Influence of the Large Scale Structures}
\label{sec:lse}

\begin{figure*}[hbtp]
	\centering
	\includegraphics[width=1\linewidth]{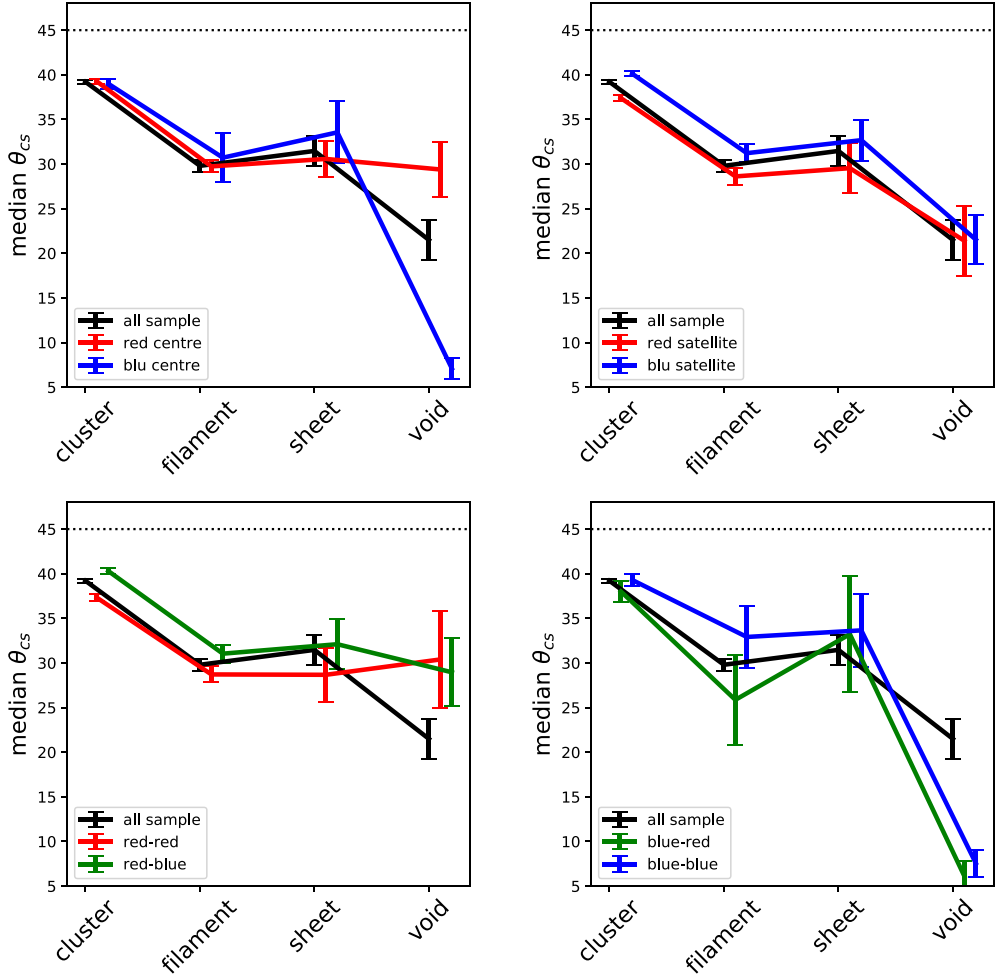}
	\caption{Average alignment angle in different environment at redshift $0$.
	Galaxy pairs are divided into different catalogues according to the color
	of central or satellite galaxies.
	The upper-left panel shows the difference between galaxy pairs with red
	and blue centrals. The upper-right panel compares the samples with red
	satellites and blue satellites. The lower left panel shows red centrals -
	red satellites samples versus red centrals - blue satellite, while the lower right shows
	blue centrals - red satellites sample versus blue centrals - blue satellites sample.
	A black solid line indicating the overall trend is shown in all panels.
	Black dotted line at $\langle\theta\rangle=45^{\circ}$ show the average
	alignment level of random distribution of satellites.
	The error bars indicate the Poisson error of mean $\theta_{CS}$.
	To make the error bars clear, we shift lines a bit horizontally.
	}
	\label{angle_lse}
\end{figure*}

\begin{table}[htpb]
	\centering
	\caption{
		The mean $\theta_{CS}$ in different environments. Sub-samples are the same
		as \reftab{Table2}. For short "satellite galaxy" is written as "sat".
	}
	\label{Table3}
	\begin{tabular}{ccccc}
		\hline
		             & Cluster              & Filament             & Sheet                & Void                 \\
		\hline
		all sample   & $39.2\pm0.2^{\circ}$ & $29.8\pm0.7^{\circ}$ & $31.5\pm1.7^{\circ}$ & $21.5\pm2.2^{\circ}$ \\
		\hline
		red central  & $39.3\pm0.3^{\circ}$ & $29.7\pm0.7^{\circ}$ & $30.6\pm2.0^{\circ}$ & $29.4\pm3.0^{\circ}$ \\
		blue central & $39.0\pm0.6^{\circ}$ & $30.7\pm2.8^{\circ}$ & $33.5\pm1.2^{\circ}$ & $7.1 \pm1.2^{\circ}$ \\
		\hline
		red sat      & $37.4\pm0.4^{\circ}$ & $28.6\pm0.9^{\circ}$ & $29.5\pm2.7^{\circ}$ & $21.4\pm4.0^{\circ}$ \\
		blue sat     & $40.1\pm0.3^{\circ}$ & $31.2\pm1.0^{\circ}$ & $32.7\pm2.3^{\circ}$ & $21.5\pm2.8^{\circ}$ \\
		\hline
		red - red    & $37.3\pm0.4^{\circ}$ & $28.7\pm0.8^{\circ}$ & $28.7\pm2.8^{\circ}$ & $30.4\pm5.3^{\circ}$ \\
		red - blue   & $40.3\pm0.3^{\circ}$ & $31.0\pm1.0^{\circ}$ & $32.1\pm2.7^{\circ}$ & $28.9\pm3.9^{\circ}$ \\
		blue - red   & $38.0\pm1.2^{\circ}$ & $25.8\pm4.7^{\circ}$ & $33.2\pm6.6^{\circ}$ & $6.1 \pm1.6^{\circ}$ \\
		blue - blue  & $39.3\pm0.6^{\circ}$ & $32.9\pm3.5^{\circ}$ & $33.6\pm4.0^{\circ}$ & $7.5 \pm1.5^{\circ}$ \\
		\hline
	\end{tabular}
\end{table}

\begin{figure}[]
	\centering
	\includegraphics[width=0.5\linewidth]{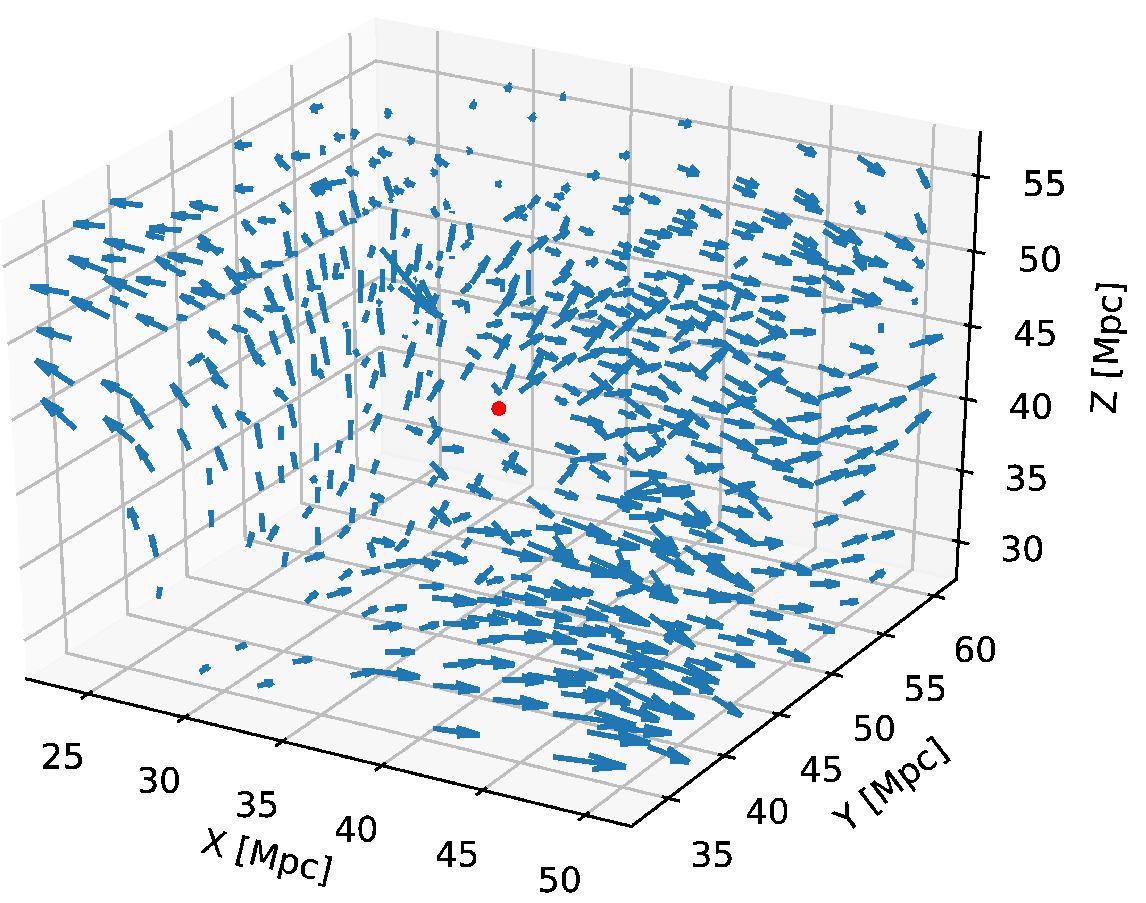}
	\quad
	\includegraphics[width=0.45\linewidth]{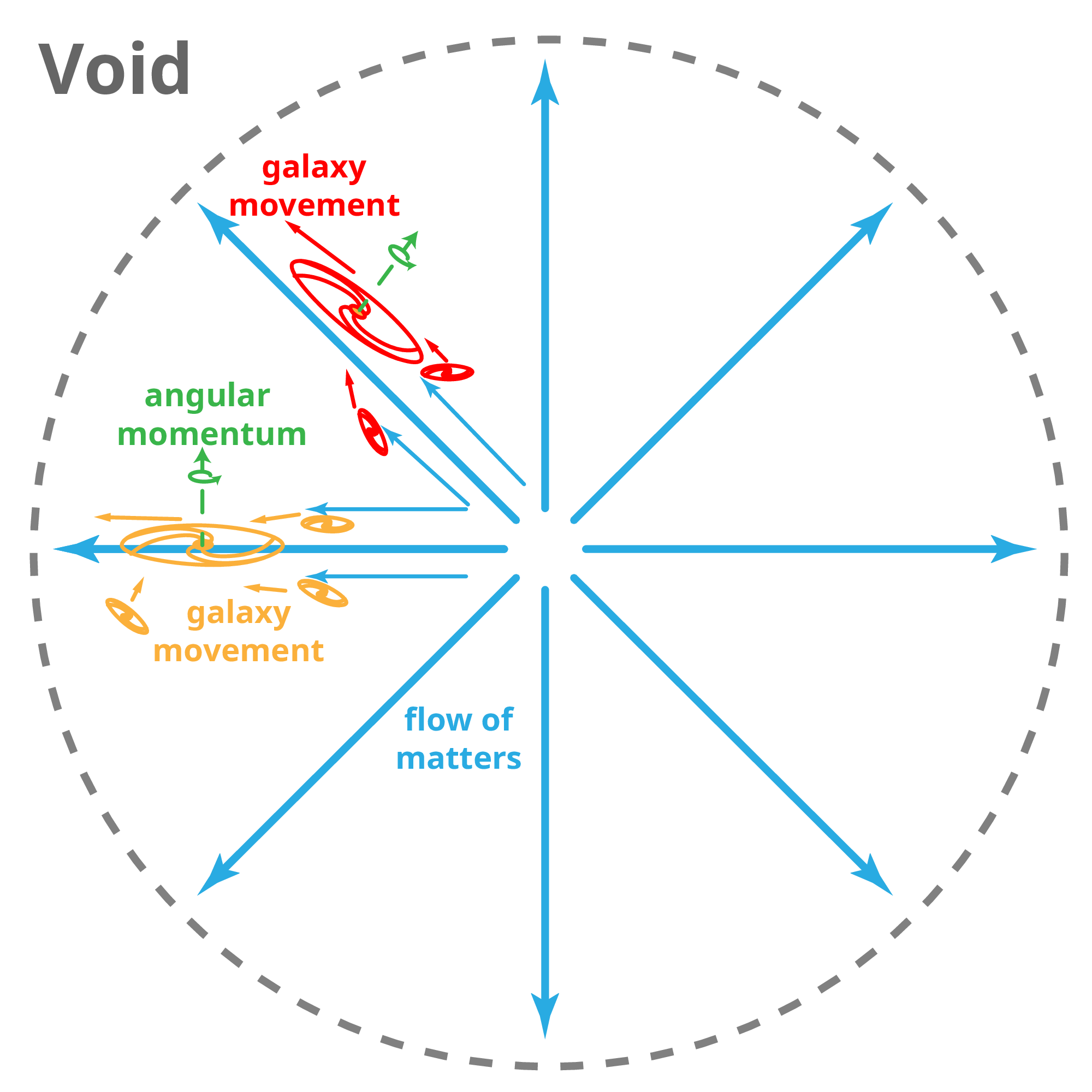}
	\includegraphics[width=0.5\linewidth]{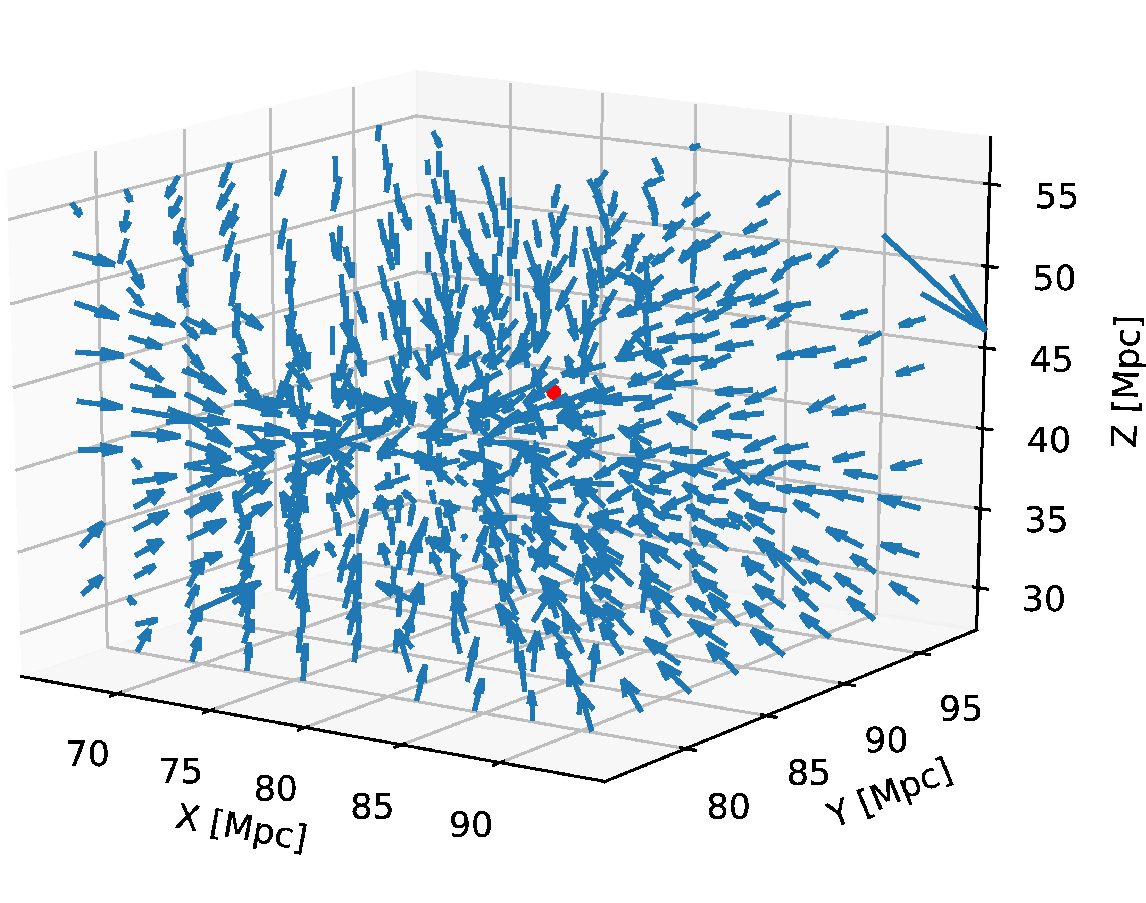}
	\quad
	\includegraphics[width=0.45\linewidth]{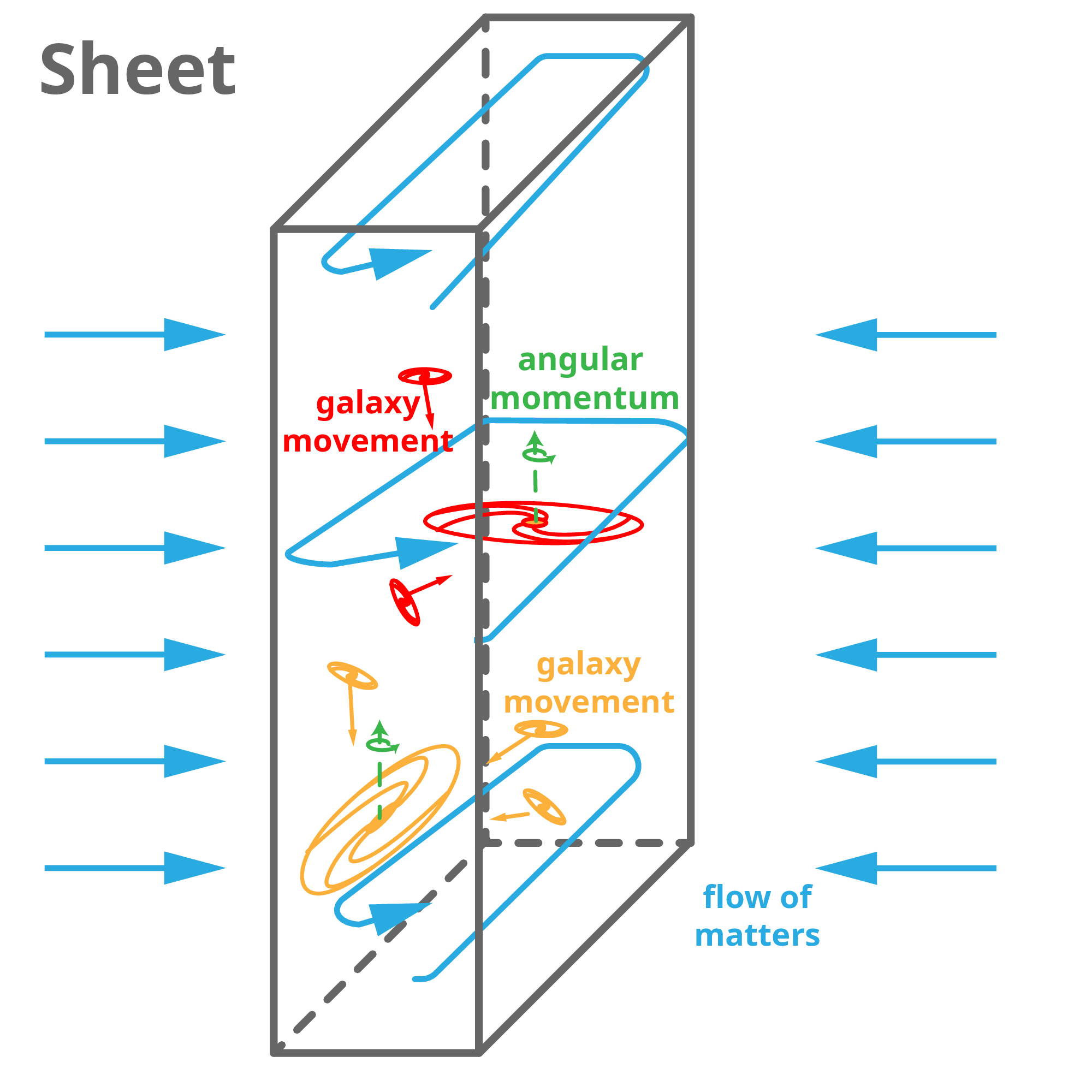}
	\includegraphics[width=0.5\linewidth]{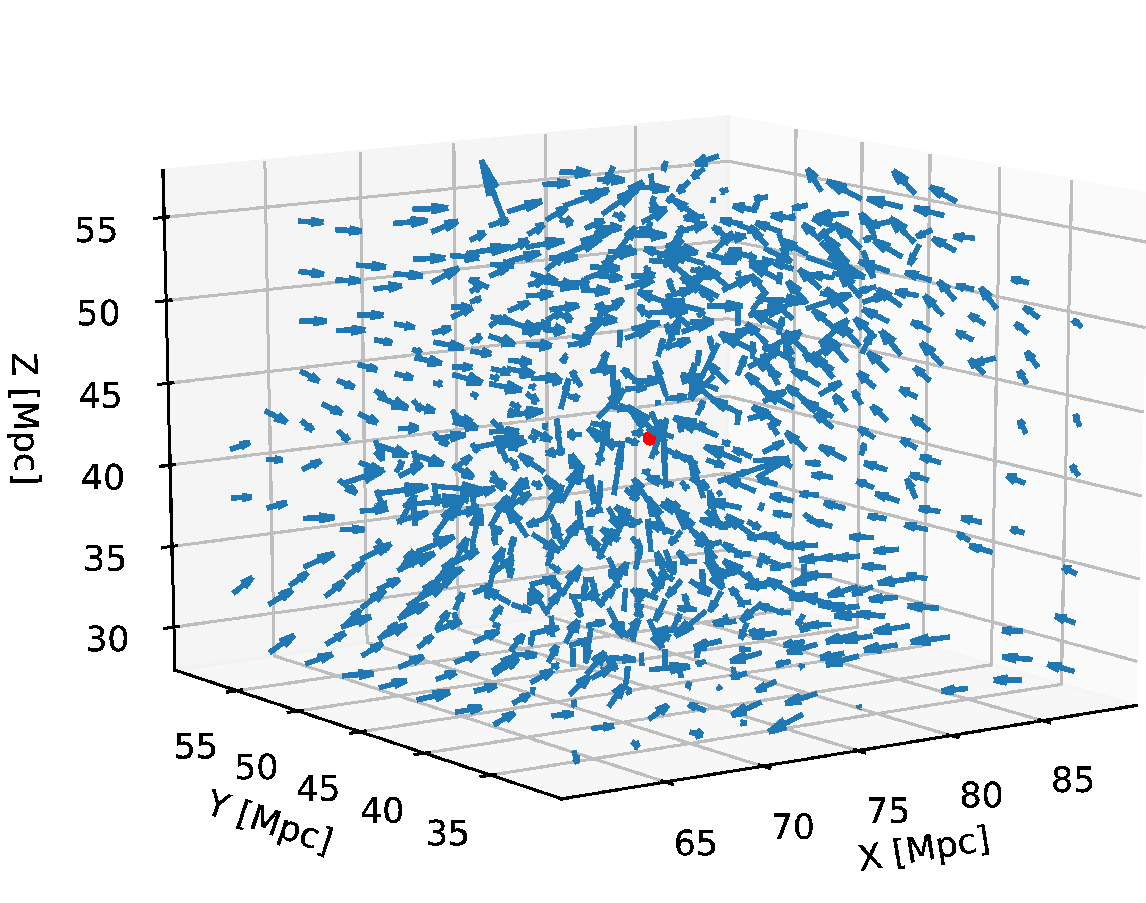}
	\quad
	\includegraphics[width=0.45\linewidth]{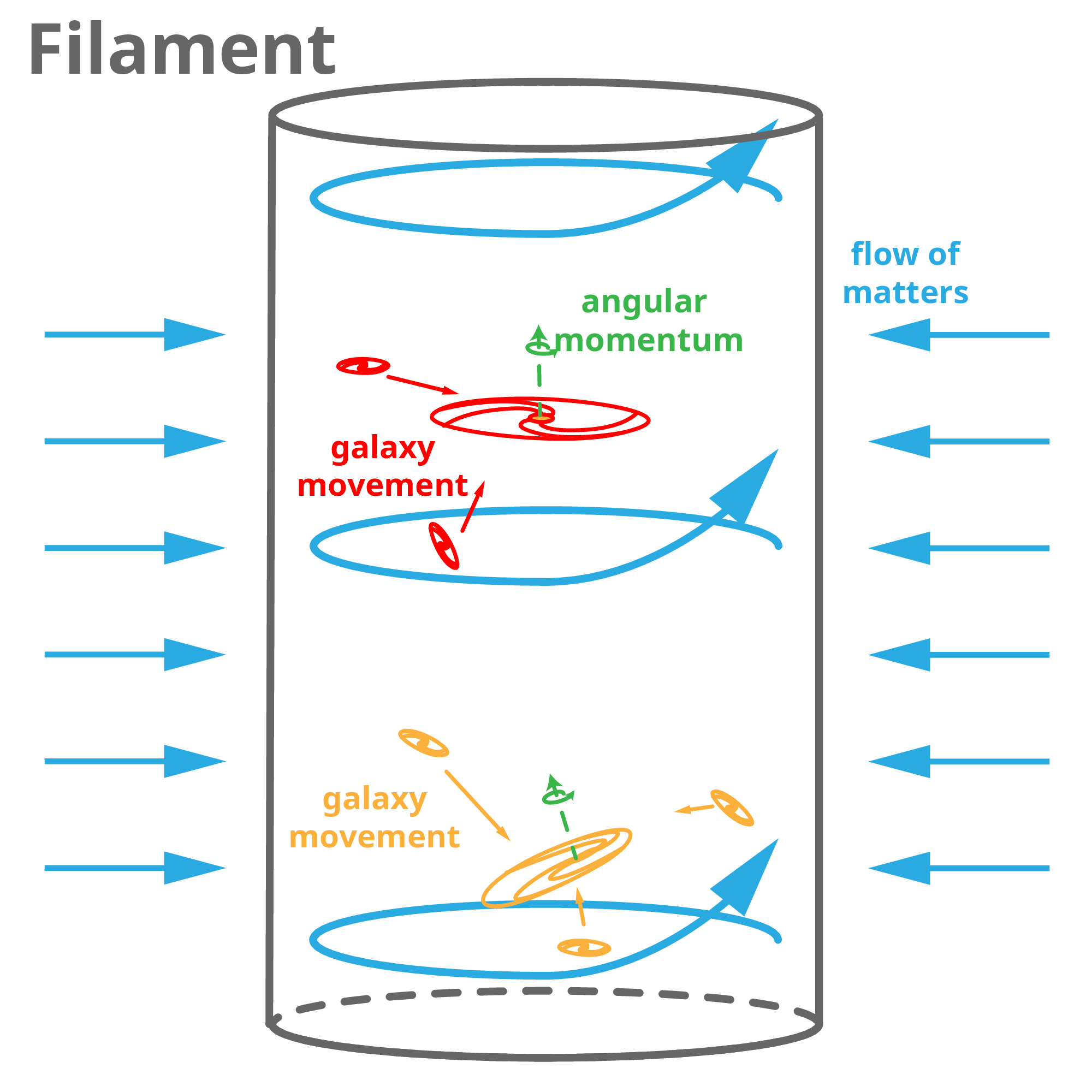}
	\caption{
		The left column show the velocity field of $10{\rm Mpc}^3$ box around
		center of void, sheet and filament. In the right column, we use cartoons to
		illustrate the movement of galaxies (red and yellow arrows) and matters in back ground(blue arrows), as well as
		the possible alignment patterns between central and satellite galaxies.
		From Top to bottom, each row shows the situation of one environments, i.e., void, sheet and filament.
		The velocity field of clusters is very messy, thus we don't show its figures here.
	}
	\label{cartoon}
\end{figure}

\begin{figure*}[hbtp]
	\centering
	\includegraphics[width=1\linewidth]{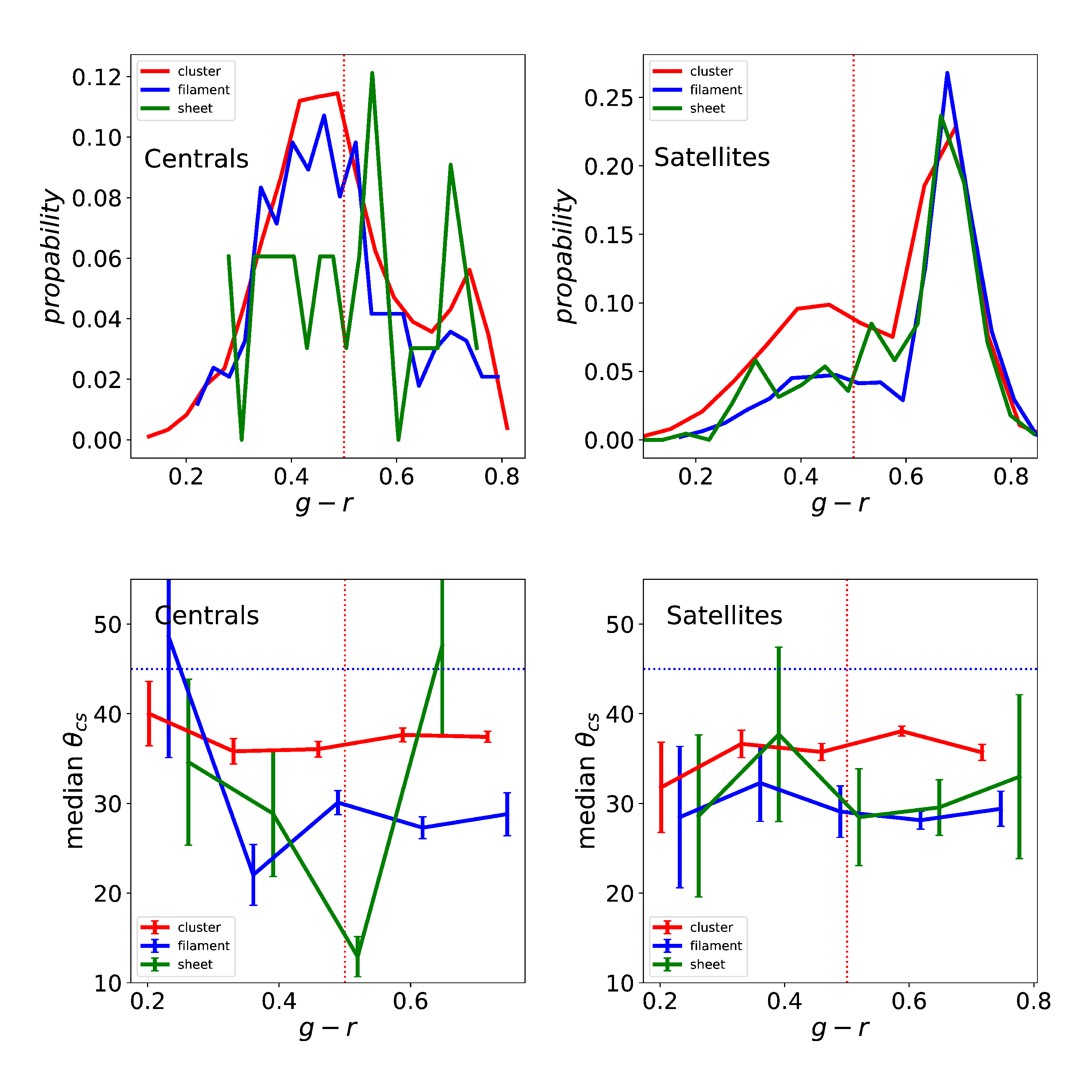}
	\caption{
		The upper two panels show the distribution of central galaxies' color (upper left) and satellite galaxies'
		color (upper right). Lower panels show the mean alignment angle $\theta_{CS}$
		as a function of central galaxies' color (lower left) and satellite galaxies'
		color (lower right). Curves for different sub-samples are distinguished by color
		as legend shows.
		Red dotted vertical line shows the division between blue and red galaxies.
		In lower panels, black dotted horizontal line shows the average alignment
		angle of random distribution.
		We don't show voids galaxies here because there are too less of them.
		To make the error bars clear, we shift lines a bit horizontally.
	}
	\label{color_web_c}
\end{figure*}

To explore the dependence of central-satellite alignment angle on the large-scale environment, we
first calculate the median angle in cluster, filament, sheet and void, which
is depicted in \reffig{angle_lse}. The exact values of mean angle are given
in \reftab{Table3}.

As shown in \reffig{angle_lse}, the alignment signal
are increasing from environments of cluster to filament, sheet, then void. This
trend is much more stronger than the results in \cite{Wang2018a}. Such environmental
dependency could be explained by tracing the accretion history in the different
large-scale environments (\citealt{Codis2012}). To illustrate this, we plot \reffig{cartoon}.

In void, the matters flows radially away from the central region to the surrounding dense region. The velocity field of void in left top subplot of
\reffig{cartoon} show this trend clearly.
During this process,
the angular momentum of the gas is thus perpendicular to the
radial direction, leading to the axis of the galaxy formed more likely aligning with the same radial direction, as represented by the right top subplot of \reffig{cartoon}.
On the other hand, the galaxies formed in voids always tend to move outward radially, on which the satellite accretions occur.
By this way, the alignment of the satellite galaxies is likely to be along the main axis of the central galaxy, producing the stronger alignment signals.  \cite{Trujillo2006}, \cite{Brunino2007} and \cite{Varela2012}  paper found that galaxies at the edge of void have a
rotating axis both parallel to the outer boundary of void and perpendicular to the radial direction, which is in agreement with our conclusion.

According to the Zeldovich's approximation, the gas accretion is always
anisotropic and the sheets usually appear firstly during the formation of
cosmic web, because there is always one direction along which the matter
collapse fastest, corresponding to the maximum eigenvalue of the deformation
matrix, so as to form a 2-dimensional pancake firstly, i.e., the sheet
structure. While accreted onto sheets from both sides, the gas almost keeps
its angular momentum parallel to the plane of sheets. Consequently, the
galaxies formed on the sheet are having their angular momentum parallel to
the sheet plane, as the middle right subplot of \reffig{cartoon} shows. At meanwhile, the major axis could be either perpendicular (red central galaxy middle right subplot of \reffig{cartoon})
or parallel (orange central galaxy) to the sheet, else lying in an intermediate state between them.
However, for the satellite galaxies, the accretion occurs in the sheet plane where the galaxies are assembled. Therefore, the position vectors of the satellite galaxies are possible to be perpendicular to the major axies of the central galaxy, causing the alignment signals to weaken and $\theta_{cs}$ to increase.

Matter in sheets would collapse further  and form filaments. Gas swirl around  and flow into filaments to form spirals that having angular momentum aligned along the
direction of filaments, as bottom subplots of \reffig{cartoon} shows. Therefor newly formed galaxies (usually blue disk
galaxies) tend to have their major axis perpendicular to filaments
(\citealt{Codis2015, Welker2018}). Since galaxies form within the cylinder of
filament, they gather together via the direction of filament.
We check the entering direction of satellites in filaments.
In our samples, $60.7\%$ of satellites have their the angles between their position vectors at entering time and the
direction of filaments smaller than $60^{\circ}$. Note that $60^{\circ}$ is
the mean value of angle separation for satellites with random distribution in 3-D space.
Thus satellites enter
their host halo in a direction that is perpendicular to the plane of center galaxy. The alignment
signal becomes weaker. However, since the shape of center galaxies make change
as they also keep accreting gas from filaments, their major axis will be
gradually redirected to the direction of filaments. On the other hand, early
accreted satellite galaxies will move closer to major axis of center galaxies
when they fall closer to them. Because of above two mechanism, old red center
galaxies consequently present stronger alignment signal (\citealt{Yang2006, Welker2018}).

In clusters, satellites galaxies are more likely come from the conjunct filaments.
However, for individual galaxy, surrounding gas is so defused with complex movement that they could be accreted
in any direction. Early formed center galaxies may have their satellites and
accreted mass via the same direction. Because at early stage of cluster  growing,
gas may flow in in the same direction as satellites. For those late
formed galaxies, position of accreted satellite could be totally irrelevant
to incoming gas.

Above descriptions are favored by many works
(e.g., \citealt{Codis2012,Codis2015, Welker2018}). They can well explain the
influence of large scale structures on galaxies' central-satellite alignments.
However, the alignment signal
in simulation is much stronger than that in observations, which indicates that the simulations might have not fully reconstruct
the non-linear process and sub-grid physics such as thermal and kinetic feedback.

When we take views from galaxy color, we found that the dependency on large scale environment
is not influenced by the color of centrals or satellites. The curves of sub-samples
have nearly the same slope as the black curve of the whole sample, while
the amplitude shift keeps almost constant in all environment. Only in
voids, the $\theta_{CS}$ of blue centrals fall rapidly while that of red
centrals become flat. This sudden change may not be reliable, since the number of galaxy
pairs in voids is much less then in other environments.

\begin{figure}[hbtp]
	\centering
	\includegraphics[width=1\linewidth]{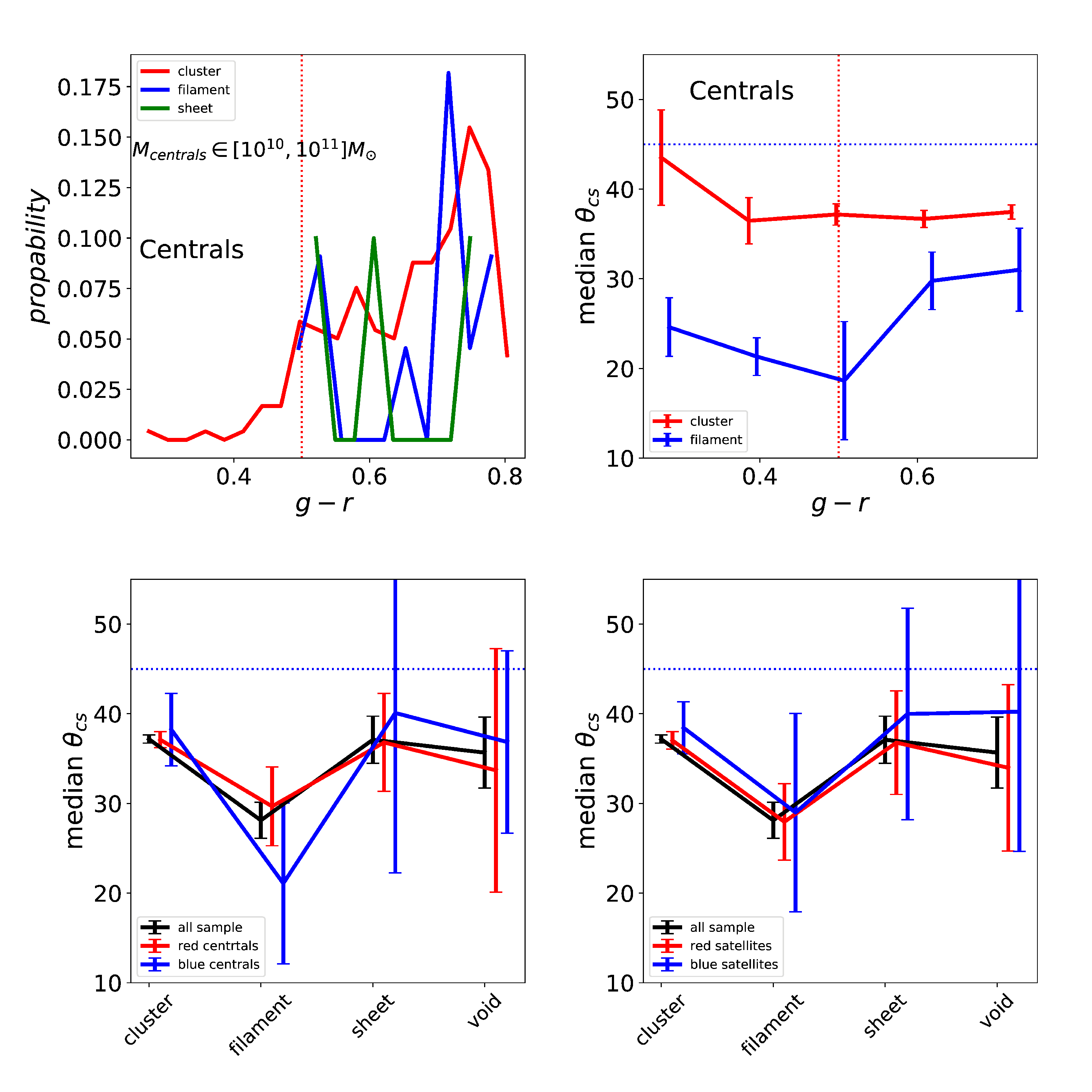}
	\caption{Top left subplot shows the color distribution of central galaxies
	with mass of $10^{10} \sim 10^{11} {\rm M_{\odot}}$. The pattern is the same
	as top left subplot of \reffig{color_web_c}. Other subplots show the
	alignment of galaxy pairs with central galaxies within the same range. Top 
	right subplot show the relation between median alignment angle and color 
	of central galaxies, which is the same as bottom left subplot in \reffig{color_web_c}.
	The curves for sheet and void galaxies are not shown because the there are
	few samples. Two bottom subplots shows the alignment-environment relation
	for galaxies with different colors, which are the same as top rows in \reffig{angle_lse}.
	}
	\label{color_web_c_part}
\end{figure}

Most parts of these $\theta_{CS}$-environment curves have the same trends
as \cite{Wang2018a}'s work. In their results, the alignment of blue centrals
is strongly dependent on environment, while red centrals does not. For the EAGLE
galaxies, blue and red centrals have same $\theta_{CS}$ in clusters, while
SDSS DR7 data shows significant discrepancy between red and blue centrals in
clusters.

We also investigated the color dependence on the environments. We found a blue
biased center galaxy group, as \reffig{color_web_c} shows. The galaxy color
distribution is almost the same in cluster, filament and sheet, except
that satellites in clusters have an extra peak in blue region. This distribution is
totally different from SDSS data (\citealt{Wang2018a}). In \cite{Wang2018a},
the probability distribution of central galaxies has its peak at the red part, and
the fraction of blue galaxies increases from clusters to sheets. In the lower panels, we found that
$\theta_{CS}$ is independent on $g-r$ in clusters. In filaments $\theta_{CS}$
decrease slightly with galaxies' color. The relation become unstable in
sheets due to the limited galaxy number. In \cite{Wang2018}, the
alignment-color relation has steady falling trend in all environment, and the
slopes are identically the same.

We have found that EAGLE galaxies do not fully recover the color distribution
as observations. However we could not say that this is the only reason for the 
inconsistences of the color dependence of alignment-environment relation between
this work and \cite{Wang2018a}.
\cite{Guo2016} claimed that, in EAGLE simulation, the passive fraction of 
galaxies with mass between $10^{10}$ to $10^{11} {\rm M_{\odot}}$ is quite
reasonable compared with observations. Thus we further check the alignment 
of galaxy pairs with central galaxies within this range. The top left subplot
of \reffig{color_web_c_part} shows that, for specific central galaxy mass range,
the color distribution is much more closer to observations (see Figure 3 in \cite{Wang2018a}).
But top right subplot tells us the $\theta_{cs}$ - $(g-r)$ relation of this sub-sample is still inconsistent
with \cite{Wang2018a}'s results. Moreover, the alignment angle is less related 
with environment for both blue and red galaxies, while in observations blue centrals have stronger alignment signal than reds (\citealt{Wang2018a}).
Such differences imply that, even for galaxy catalogue with right color
distribution, the alignment - color relation is not well recovered due to 
some sophisticated sub-grid physics.

\subsection{LSE at High Redshift}

\begin{figure}[hbtp]
	\centering
	\includegraphics[width=1\linewidth]{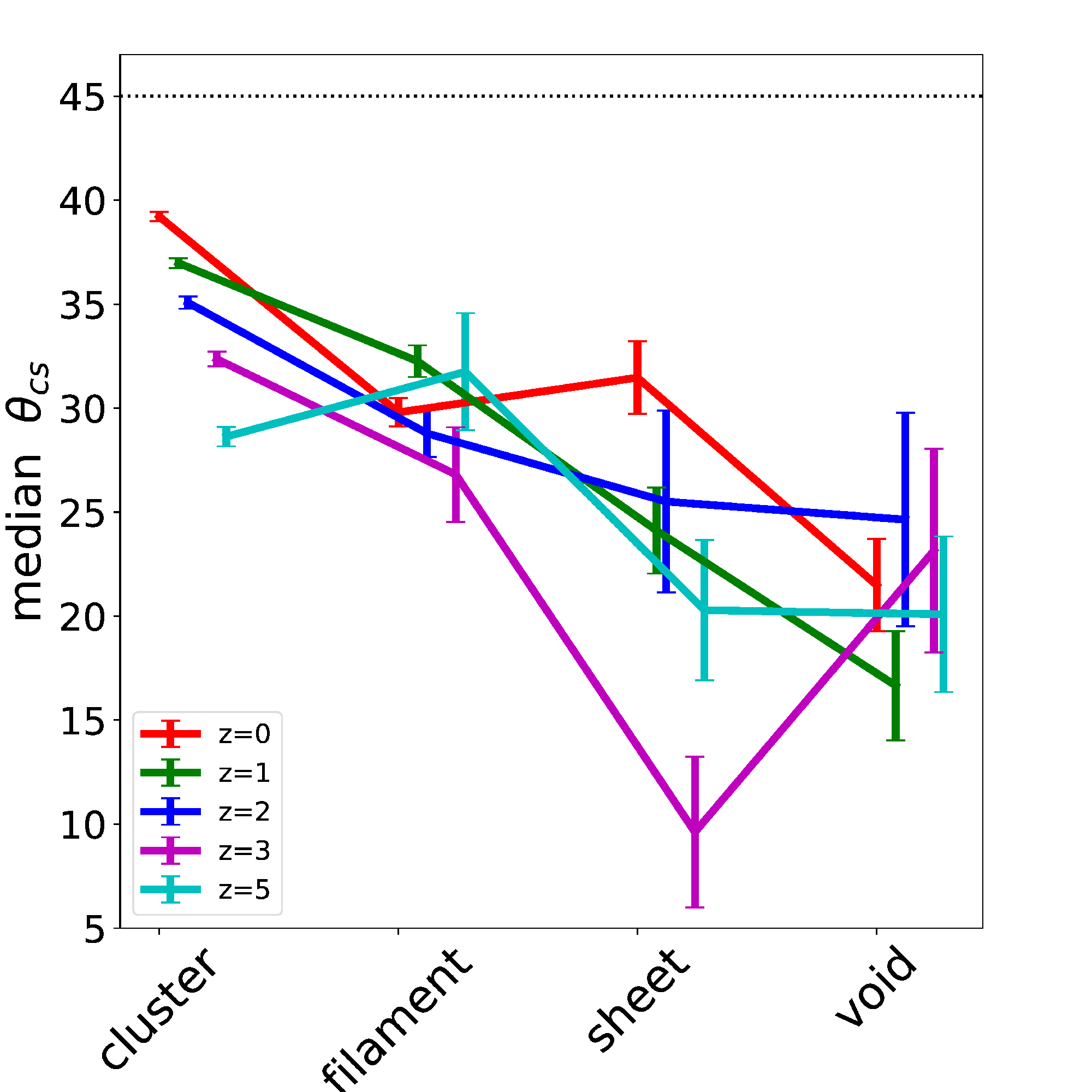}
	\caption{Alignment angle distribution in different environments at different redshifts. Alignment-environment relation curves for different
		redshifts are distinguished by different color as legend shows. Black dashed
		line at $\theta_{CS}$ indicates the alignment angle of random distribution.
		To make the error bars clear, we shift lines a bit horizontally.
	}
	\label{angle_lse_z}
\end{figure}

\begin{figure}[hbtp]
	\centering
	\includegraphics[width=1\linewidth]{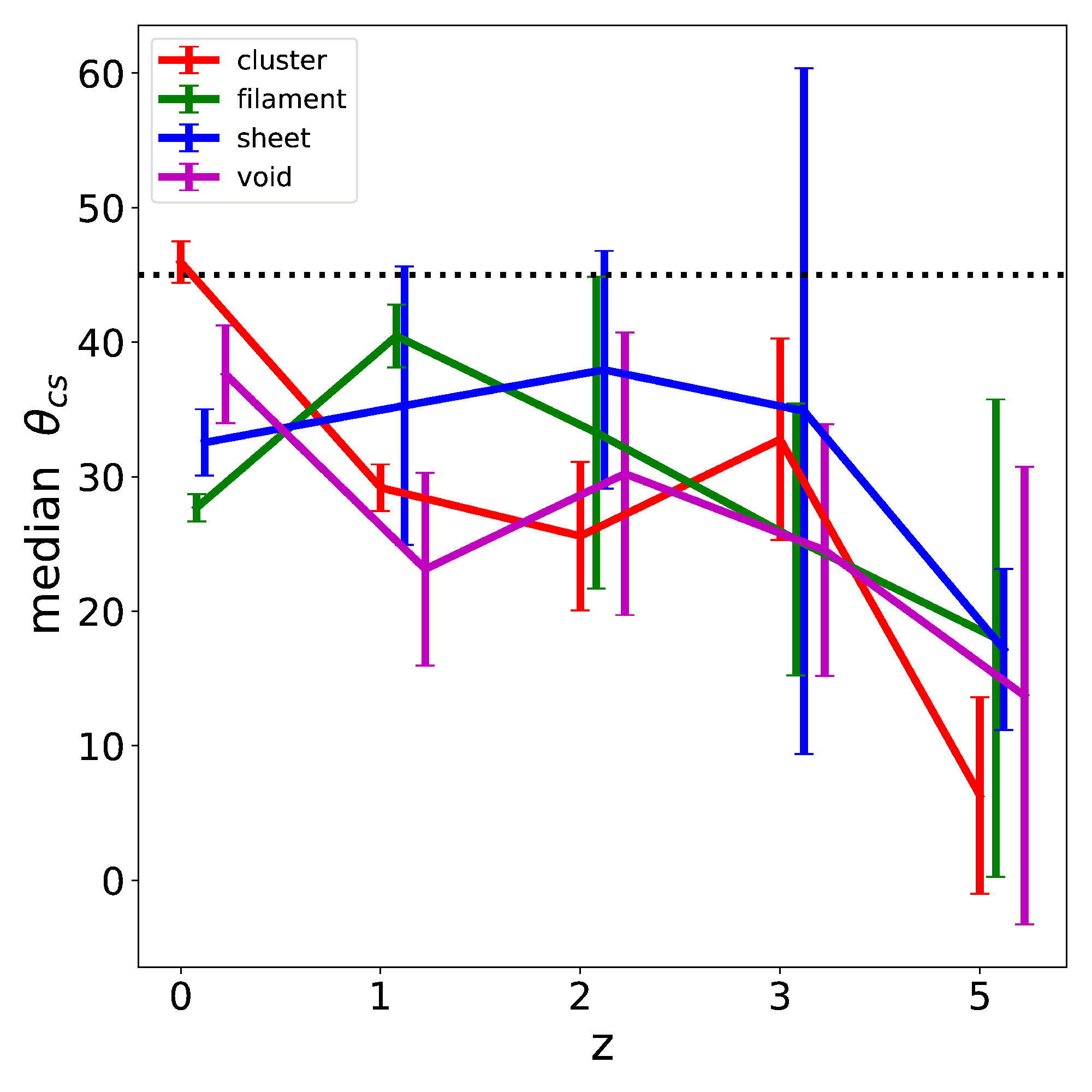}
	\caption{The change of mean alignment angle of galaxies in four dark matter
		halos. Four halos are selected from different large scale structures, i.e.,
		cluster, filament, sheet and void. Each line shows the statistic of galaxy
		alignment between one center galaxy and its satellite galaxies.
		To make the error bars clear, we shift lines a bit horizontally.
	}
	\label{angle_z_group}
\end{figure}

In order to investigate the evolution of the alignment angle, we also calculate
the alignment angle at higher redshifts, e.g. $z=1,2,3,5$, \reffig{angle_lse_z} illustrates
the probability distribution of the alignment angle at those redshifts.
We found that the alignment-environment correlation exists at all redshifts.
The strength of the alignment signal increases from clusters, filaments to
sheets then voids. The slopes of those curves at different redshifts are quite close,
implying that the difference between structures does not change significantly for a long time.
As we have mentioned in section \ref{sec:lse}, the alignment comes from the different
mass accretion paths both of central and satellite galaxies.
A reasonable scenario is that the mass accretion follows the different paths
in the different large scale environments, resulting in the environmental dependency
of the alignment. Thus we would expect environmental dependency of alignment
at any time that large scale structures exists.

In clusters, central-satellite galaxy pairs show stronger alignment
signal at higher redshifts.
This evolution trend still exists in filaments, but become merely observed
in sheets and voids.
It is found that early accreted satellites, in another words, satellites closer
to the center galaxies, prefer to align along the galaxies' major axis
(\citealt{Yang2006, Faltenbacher2008, Wang2018}) . One
possible explanation is that the satellites move closer to the major axis after
falling into their host halo (\citealt{Welker2018}) . However our \reffig{angle_lse_z} implies another
possible routine: early accreted satellites tend to enter their host halos in paths that are closer to
the major axis than those entering late. Late accreted satellites dilute alignment signal resulting in increasing average $\theta_{CS}$.

To confirm which process dominates the evolution of central-satellite alignments, we choose 4 central galaxies to trace how the alignment
angles of their satellites evolves with redshifts. These four center galaxies are deliberately chosen
from the four different structure types. For each central galaxy, we calculate the mean $\theta_{CS}$ of its satellites at each snapshots, then
plot $\langle\theta_{CS}\rangle$ versus $z$ in \reffig{angle_z_group}. In
\reffig{angle_z_group}, $\langle\theta_{CS}\rangle$ of cluster galaxies decreases
from $z=0$ to $z=5$. It supports our previous assumption that the early accreted satellites was accreted closer to the major axis of central galaxy than the late accreted ones. The filament, sheet and void galaxies do not have monotonic evolution change curves. That is why we
did not observe alignment angle weakening with time in other structures except
clusters.


\section{Conclusion and Discussion}
\label{sec:con}

Using data from EAGLE hydrodynamic simulation, we explore the large-scale
environment dependence of the alignment angle between central galaxy major
axis and satellite galaxy position vector. This is mainly a follow up work
of \cite{Wang2018a}, thus all results are compared with their results based on
SDSS observation data. The general conclusions are
summarized below:
\begin{itemize}
	\item In consistent with the results in \cite{Wang2018a},
	      the alignment signal between the major axis of central galaxies and the position vector of satellite galaxies in EAGLE simulations exhibits environmental dependence. Average         \
	      alignment angle
	      decreases gradually when the environment changes from cluster to filament, sheet and void.
	      However, the amplitude of alignment signal in simulation, as well as the
	      environmental dependency, are much stronger than {results extracted from }observations. Further
	      improvements on the sub-physics of simulation may overcome this discrepancy. It is also possible that observational contamination
	      dilute the alignment signal.

	\item We found that EAGLE galaxies don't recover the dependency of alignment
		  on galaxy color. The fact that EAGLE produces more blue central galaxies 
		  than observations takes part of emergence of this bias. However, the
		  trends are not right even for sub-samples with right color distribution. It is possible that the color of both red and blue galaxies is wrongly assigned. 
		  To testify it, a through investigation into sub-grid physics 
		  is required in future works.

	\item We found that the influence of large scale structures exists at high
	      redshifts. It demonstrates that the phenomenon of alignments of satellite
	      galaxies is mainly dynamically driven process which is  largely determined by the
	      flows of matters.
\end{itemize}

Comparing with the alignment signal extracted from observations
in \cite{Wang2018a}, there are main two differences in our results: the alignment signal is too strong,
and galaxy colors are mis-assigned. However, it is still worthwhile to consider
the large scale structure effect on alignment within the scope of simulations. The
simulation basically reproduce the trends for overall alignment signal as well as
for the dependency on large scale environment. Since the dynamic processes of
simulations is promising, we could explore the inner driven for large scale
structure dependency. And we confirm that the alignment signal reported in
this work and \cite{Wang2018a} can be explained by the matter accretion
scenario proposed in previous works \cite{Codis2015,Codis2012, Welker2018}.

\normalem
\begin{acknowledgements}

	We thanks referee for through comments and suggestions to this paper.
	YW is supported by NSFC No.11803095, YW and ZMG are supported by NSFC No.11733010.
	ZMG thanks Weishan Zhu for helpful discussions and suggestions.  We acknowledge the Virgo Consortium
	for making their simulation data available. The eagle simulations were performed using the DiRAC-2 facility at
	Durham, managed by the ICC, and the PRACE facility Curie based in France at
	TGCC, CEA, Bruyèresle-Châtel.

\end{acknowledgements}

\bibliographystyle{raa}
\bibliography{D:/OneDrive/Documents/Work/Bib/Lib}

\begin{thebibliography}{54}
\providecommand\natexlab[1]{#1}
\providecommand\JournalTitle[1]{#1}

\bibitem[Agustsson \& Brainerd(2006{\natexlab{a}})]{Agustsson2006a}
Agustsson, I., \& Brainerd, T.~G. 2006{\natexlab{a}}, \apj, 650, 550

\bibitem[Agustsson \& Brainerd(2006{\natexlab{b}})]{Agustsson2006b}
Agustsson, I., \& Brainerd, T.~G. 2006{\natexlab{b}}, \apj, 644, L25

\bibitem[Agustsson \& Brainerd(2010)]{Agustsson2010}
Agustsson, I., \& Brainerd, T. T.~G. 2010, \apj, 709, 1321

\bibitem[Arag{\'o}n-Calvo {et~al.}(2010)]{Aragon-Calvo2010}
Arag{\'o}n-Calvo, M.~A., van~de Weygaert, R., \& Jones, B. J.~T. 2010, \mnras,
  408, 2163

\bibitem[Arieli {et~al.}(2010)]{Arieli2010}
Arieli, Y., Rephaeli, Y., \& Norman, M.~L. 2010, \apj, 716, 918

\bibitem[Aubert {et~al.}(2004)]{Aubert2004}
Aubert, D., Pichon, C., \& Colombi, S. 2004, \mnras, 352, 376

\bibitem[Bagla \& Prasad(2006)]{Bagla2006}
Bagla, J.~S., \& Prasad, J. 2006, \mnras, 370, 993

\bibitem[Baldry {et~al.}(2004)]{Baldry2004b}
Baldry, I.~K., Balogh, M.~L., Bower, R., Glazebrook, K., \& Nichol, R.~C. 2004,
  in American Institute of Physics Conference Series, Vol. 743, The New
  Cosmology: Conference on Strings and Cosmology, ed. R.~E. {Allen}, D.~V.
  {Nanopoulos}, \& C.~N. {Pope}, 106

\bibitem[Bond {et~al.}(1996)]{Bond1996}
Bond, J.~R., Kofman, L., \& Pogosyan, D. 1996, \nat, 380, 603

\bibitem[Brainerd(2005)]{Brainerd2005}
Brainerd, T. 2005, 628, L101

\bibitem[Brunino {et~al.}(2007)]{Brunino2007}
Brunino, R., Trujillo, I., Pearce, F.~R., \& Thomas, P.~A. 2007, \mnras, 375,
  184

\bibitem[Codis {et~al.}(2018)]{Codis2018a}
Codis, S., Jindal, A., Chisari, N.~E., {et~al.} 2018, \mnras, 481, 4753

\bibitem[Codis {et~al.}(2012)]{Codis2012}
Codis, S., Pichon, C., Devriendt, J., {et~al.} 2012, \mnras, 427, 3320

\bibitem[Codis {et~al.}(2015)]{Codis2015}
Codis, S., Pichon, C., \& Pogosyan, D. 2015, \mnras, 452, 3369

\bibitem[Davis {et~al.}(1985)]{Davis1985}
Davis, M., Efstathiou, G., Frenk, C.~S., \& White, S. D.~M. 1985, \apj, 292,
  371

\bibitem[Dolag {et~al.}(2009)]{Dolag2009}
Dolag, K., Borgani, S., Murante, G., \& Springel, V. 2009,
  arXiv:arXiv:0808.3401v2

\bibitem[Dong {et~al.}(2014)]{Dong2014}
Dong, X.~C., Lin, W.~P., Kang, X., {et~al.} 2014, \apjl, 791, L33

\bibitem[Faltenbacher {et~al.}(2008)]{Faltenbacher2008}
Faltenbacher, A., Jing, Y.~P., Li, C., {et~al.} 2008, The Astrophysical
  Journal, 675, 146

\bibitem[Faltenbacher {et~al.}(2009)]{Faltenbacher2009}
Faltenbacher, A., Li, C., White, S. D.~M., {et~al.} 2009, Research in Astronomy
  and Astrophysics, 9, 41

\bibitem[Forero-romero {et~al.}(2014)]{Forero2014}
Forero-romero, J.~E., Contreras, S., Padilla, N., \& Jun, C.~O. 2014, 000,
  arXiv:arXiv:1406.0508v1

\bibitem[Forero-Romero {et~al.}(2009)]{Forero-Romero2009}
Forero-Romero, J.~E., Hoffman, Y., Gottl{\"o}ber, S., Klypin, A., \& Yepes, G.
  2009, \mnras, 396, 1815

\bibitem[Guo {et~al.}(2016)]{Guo2016}
Guo, Q., Gonzalez-Perez, V., Guo, Q., {et~al.} 2016, \mnras, 461, 3457

\bibitem[Hahn {et~al.}(2007)]{Hahn2007}
Hahn, O., Porciani, C., Carollo, C.~M., \& Dekel, A. 2007, \mnras, 375, 489

\bibitem[Holmberg(1969)]{Holmberg1969}
Holmberg, E. 1969, Arkiv for Astronnomi, 5, 305

\bibitem[Jing \& Suto(2002)]{Jing2002}
Jing, Y., \& Suto, Y. 2002, arXiv:0202064v5

\bibitem[Joachimi {et~al.}(2015)]{Joachimi2015}
Joachimi, B., Cacciato, M., Kitching, T.~D., {et~al.} 2015, \ssr, 193, 1

\bibitem[J{\~o}eveer {et~al.}(1978)]{Joeveer1978}
J{\~o}eveer, M., Einasto, J., \& Tago, E. 1978, \mnras, 185, 357

\bibitem[Kang {et~al.}(2007)]{Kang2007}
Kang, X., van~den Bosch, F., Yang, X., {et~al.} 2007, Not., 378, 1531

\bibitem[Kiessling {et~al.}(2015)]{Kiessling2015}
Kiessling, A., Cacciato, M., Joachimi, B., {et~al.} 2015, \ssr, 193, 67

\bibitem[Kirk {et~al.}(2015)]{Kirk2015}
Kirk, D., Brown, M.~L., Hoekstra, H., {et~al.} 2015, \ssr, 193, 139

\bibitem[Libeskind {et~al.}(2007)]{Libeskind2007}
Libeskind, N., Cole, S., Frenk, C., Okamoto, T., \& Jenkins, A. 2007, 374, 16

\bibitem[Libeskind {et~al.}(2005)]{Libeskind2005}
Libeskind, N., Frenk, C., Cole, S., {et~al.} 2005, 363, 146

\bibitem[Libeskind {et~al.}(2015)]{Libeskind2015}
Libeskind, N.~I., Hoffman, Y., Tully, R.~B., {et~al.} 2015, \mnras, 452, 1052

\bibitem[Libeskind {et~al.}(2014)]{Libeskind2014}
Libeskind, N.~I., Knebe, A., Hoffman, Y., \& Gottl{\"o}ber, S. 2014, \mnras,
  443, 1274

\bibitem[Sastry(1968)]{Sastry1968}
Sastry, G.~N. 1968, \pasp, 80, 252

\bibitem[Sch{\"a}fer(2009)]{Schaefer2009d}
Sch{\"a}fer, B.~M. 2009, International Journal of Modern Physics D, 18, 173

\bibitem[Schaye {et~al.}(2015)]{Schaye2015}
Schaye, J., {et~al.} 2015, \mnras, 446, 521

\bibitem[Springel(2005)]{Springel2005}
Springel, V. 2005, Notes, 1

\bibitem[Springel {et~al.}(2001)]{Springel2001}
Springel, V., White, S. D.~M., Tormen, G., \& Kauffmann, G. 2001, \mnras, 328,
  726

\bibitem[Tempel {et~al.}(2015)]{Tempel2015a}
Tempel, E., Guo, Q., Kipper, R., \& Libeskind, N.~I. 2015, \mnras, 450, 2727

\bibitem[Tempel \& Libeskind(2013)]{Tempel2013}
Tempel, E., \& Libeskind, N.~I. 2013, \aj, 775, L42

\bibitem[Trayford {et~al.}(2015)]{Trayford2015}
Trayford, J.~W., Theuns, T., Bower, R.~G., {et~al.} 2015, \mnras, 452, 2879

\bibitem[Trujillo {et~al.}(2006)]{Trujillo2006}
Trujillo, I., Carretero, C., \& Patiri, S.~G. 2006, \apjl, 640, L111

\bibitem[Varela {et~al.}(2012)]{Varela2012}
Varela, J., Betancort-Rijo, J., Trujillo, I., \& Ricciardelli, E. 2012, \apj,
  744, 82

\bibitem[Wang \& Kang(2018)]{Wang2018}
Wang, P., \& Kang, X. 2018, \mnras, 473, 1562

\bibitem[Wang {et~al.}(2018)]{Wang2018a}
Wang, P., Luo, Y., Kang, X., {et~al.} 2018, \apj, 859, 115

\bibitem[Wang {et~al.}(2014)]{Wang2014b}
Wang, Y.~O., Lin, W., Kang, X., {et~al.} 2014, \apj, 786, 8

\bibitem[Welker {et~al.}(2018)]{Welker2018}
Welker, C., Dubois, Y., Pichon, C., Devriendt, J., \& Chisari, N.~E. 2018,
  \aap, 613, A4

\bibitem[White \& Rees(1978)]{White1978}
White, S. D.~M., \& Rees, M.~J. 1978, \mnras, 183, 341

\bibitem[Yang {et~al.}(2006)]{Yang2006}
Yang, X., van~den Bosch, F. F.~C., Mo, H.~J., {et~al.} 2006, \mnras, 369, 1293

\bibitem[Zhang {et~al.}(2009)]{zhang2009}
Zhang, Y., Yang, X., Faltenbacher, A., {et~al.} 2009, \apj, 706, 747

\bibitem[Zhang {et~al.}(2015)]{Zhang2015}
Zhang, Y., Yang, X., Wang, H., {et~al.} 2015, Astrophys. J., 798, 17

\bibitem[Zhang {et~al.}(2013)]{Zhang2013}
Zhang, Y., Yang, X., Wang, H., {et~al.} 2013, Astrophys. J., 779, 160

\bibitem[Zhu \& Feng(2017)]{Zhu2017}
Zhu, W., \& Feng, L.-L. 2017, \apj, 838, 21

\end{thebibliography}
\label{lastpage}
\end{document}